\documentstyle[12pt,amssymb]{article}
\topmargin = - 1.cm
\hoffset = - 1.cm
\begin {document}
\large
\begin {center}
{\bf Analysis of the probability distribution of photocount number
of the onemode stochastic radiation}
\medskip

Virchenko Yu.P., Vitokhina N.N.
\end {center}

\begin {center}
{\bf 1. Introduction}
\end {center}

For the description of the registration process of
low-intensive electromagnetic
radiation by counters of photons, it is necessary to take
into account quantum effects. Thus, even in the case when the electromagnetic
field does not contain any noise component, i.e. its amplitude
represents {\it the pure quantum state}, the number ${\tilde n}$ of registered
photons is random and there exists the problem of determination of the
probability distribution of this random variable.
This problem becomes very complicated if the registered electromagnetic field
is the mixture of deterministic field and a noise. In this case, its state is
statistically {\it mixed} and, therefore, it is necessary to use
the appropriate density matrix. From theoretical point of view, the complexity
of the description of such a state consists of the adequate choice of
mathematical model of the electromagnetic noise and of the construction on its
basis the appropriate density matrix. Simplification of this problem arises at
sufficiently large typical frequencies of the electromagnetic field.
In this case, one may use the quasi-classical approximation as it is shown in
\cite{1}. Therefore, probabilities of registered photon numbers are
calculated on the basis of the classical (not quantum) probability distribution
$$ P_n \equiv {\rm Pr}
\{{\tilde n} = n \} = \frac {1} {n!} {\sf E} {\tilde J}^n
\exp [-{\tilde J}], \eqno (1) $$
which represents the so-called {\it composite Poisson distribution}.
It is referred to the Mandel distribution in quantum optics \cite{2}.
Here, ${\tilde J}$ is the random variable representing the energy of
electromagnetic field absorbed during the registration time $T$.

Let us consider the model of the quantum photocounter \cite {1}
of the onemode electromagnetic radiation being completely noise.
In this case, the appropriate model of the electromagnetic noise is
the complex Ornstein-Uhlenbeck process as it is proposed in the
photodetection theory. Then the random variable $\tilde J$ is represented by
the formula \cite {1}
$$\tilde J \equiv J [\tilde \zeta] =
\int\limits_0^T \left |\tilde \zeta (s) \right|^2 ds \, $$
where
$\tilde \zeta (s) = \tilde \xi (s) + i \tilde \eta (s) $, $s \in {\Bbb R} $
are trajectories of the complex process connected with real
Ornstein-Uhlenbeck's processes $\tilde \xi =
\{\xi (t); t \in {\Bbb R}\}$, $\tilde \eta =
\{\eta (t); t \in {\Bbb R} \}$ being
stochastically equivalent and independent. From the physical point of view,
they correspond accordingly to electric and magnetic constituents
of the noise electromagnetic field.

Ornstein-Uhlenbeck's processes are markovian and gaussian and they
are completely characterized by  these properties and their stationary
condition. This class of processes is parametrizated by two numbers $\nu
> 0$, $\sigma > 0$. Each Ornstein-Uhlenbeck process is completely
determined by the following formula of the conditional probability density
$w(x_0,t_0|x, t)$ of the transition from the point $x_0 \in {\Bbb R}$ at
$t_0\in {\Bbb R}$ to the point $x \in {\Bbb R}$ at $t \in {\Bbb R}$. It
depends on parameters $\nu, \sigma$ and has the following form \cite{3}
$$w (x_0, t_0 | x, t) = \left(\frac \nu {\pi \sigma \left(1-e^{-2
\nu |t - t_0|}\right)} \right)^{1/2}\ \exp\left(-\frac
{\nu\left [x - x_0 e^{-\nu |t - t_0|} \right]^2} {\sigma \left (1-
e^{-2 \nu |t - t_0|}\right)} \right) \,. \eqno (2) $$
Thus, the one-point distribution density $w(x)$, $x \in {\Bbb R}$ of the
process is determined by the formula
$$w (x) = \lim_{t_0 \to-\infty} w(x_0, t_0|x, t) =
\left (\frac \nu {\pi\sigma} \right)^{1/2} \exp\left
(-\frac {\nu x^2} \sigma \right) \,.\eqno (3) $$

The characteristic function $Q(-i\lambda)$, $\lambda \in {\Bbb R}$ of the
random variable
$J[{\tilde \xi}]$ of the process $\tilde \xi$ is given
by the known Ziegert formula \cite {4},
$$Q_{{\tilde \xi}} (\lambda) = {\sf E} \exp (-\lambda J [{\tilde
\xi}]) = \left (\frac {4 r \nu \exp (\nu T)} {(r + \nu)^2 \exp (r T) - (r-
\nu)^2 \exp (-rT)} \right)^{1/2} \,, \eqno (4)$$
where $r = \sqrt {\nu^2 + 2 \lambda \sigma} $.

Since processes $\{\tilde \xi (t) \}, \{\tilde \eta (t) \} $
are independent and equivalent, the generating function of the
random variable $J [{\tilde \zeta}] $ is found on the basis of equalities
$$ Q (\lambda) = Q_{{\tilde \xi}} (\lambda)
Q_{{\tilde \eta}} (\lambda) = Q^2_{{\tilde \xi}}
(\lambda)\,. $$

From here, we see that the Mandel distribution $P_n $
determined by the formula (1) and by the probability distribution
of the random variable ${\tilde J}$ induced by the probability distribution
of the process $\tilde \zeta =
\{\zeta (t); t \in {\Bbb R} \}$ are very complex.
It is easy to obtain the asymptotic formula of the probability
distribution $P_n$ at $T \to 0$. It has the form $P_n/P_n^{(0)} \to 1$
where $P_n^{(0)}$ is the following Poisson distribution \cite {1}
$$P_n^{(0)} = \frac 1 {n!}
\left (\frac {\sigma T} \nu\right)^n \exp\left (-\frac {\sigma
T} \nu\right) \,.$$

The purpose of the present work is the construction of the convenient
calculation algorithm giving probabilities $P_n$ with
the guaranteed accuracy at the sufficiently small value of $T$ on the basis of
formulas (1)-(4).
Our approximations are based on the simple idea of the series expansion
of the Mandel distribution, i.e. we represent it in the form
$$P_n = \frac 1 {n!} \sum_{l=0}^\infty \frac {(-1)^l}{l!}
{\sf E} {\tilde J}^{n+l} \,. \eqno (5)$$
It is easy to show that each moment ${\sf E} {\tilde J}^n$, $n\in {\Bbb N}$
is proportional to $T^n$ at $T \to 0$. Then one may expect that just such an
expansion is appropriate for the solution of the above-mentioned problem
at small values of $T$.
On this way, we should solve two questions. The first, it is necessary
to give the general foreseeable formula of moments ${\sf E} {\tilde J}^n$
depending on the number $n$ as on the parameter.
It appears that it is rather problematic to obtain such a formula.
Instead of this, it is possible, however, to point out the specific algorithm
of the consecutive calculation of these values. One may think that
this algorithm is the solution of the first problem.
The second question consists of the estimation of the series remainder
$$\sum_{l=N+1}^\infty \frac 1{l!}{\sf E} {\tilde J}^{n+l} \eqno (6)$$
for any $N \in {\Bbb N}$. The presence of such an estimation permits
to give the guaranteed accuracy of the approximation $P_n^{(N)}$
based on the account only of first $N $ terms in the series (5).
Below, we solve these two problems.

\begin {center}
{\bf 2. Sequence algebra of power series coefficients}
\end {center}

Let us introduce into consideration the linear manifold
${\frak L}_f = \{f (z) \}$ of functions depending on the variable
$z \in {\Bbb C}$. Each of them is analytic in the appropriate circle having
nonzero radius and the centre in the point 0. Thus, each function
of ${\frak L}_f $ is presented as the series
$$ f (z) = \sum_{k=0}^\infty a_k z^k \eqno (7)$$
and each such a series is completely determined by the sequence of
coefficients $\langle a_k \in {\Bbb C}; k \in {\Bbb N}_+ \rangle$ which
is regarded as the infinite ordered collection.
In this connection, we shall consider also the linear
manifold $ {\frak L} = \{A \}$ of coefficient sequences
$A = \langle a_k; k\in {\Bbb N}_+ \rangle$ corresponding to serieses (7).
We shall name elements of such sequences as {\it components} also.
Linear operations on manifold ${\frak L}$ are introduced by the natural way.
Namely, let $A = \langle a_k; k\in {\Bbb N}_+ \rangle $
and $B =\langle b_k; k \in {\Bbb N}_+ \rangle $ be two sequences.
Then sequences $A + B$ and $\lambda A$ for any
$\lambda \in {\Bbb C}$ are determined by formulas
$$A+B = \langle a_k + b_k; k\in {\Bbb N}_+ \rangle \, \quad
\lambda A = \langle \lambda a_k; k \in {\Bbb N}_+ \rangle \,.$$

There exists the natural one-to-one correspondence between introduced
manifolds
${\frak L} \, \leftrightarrow \, { \frak L}_f$ determined by the formula (7).
We designate the mapping ${\frak L} \mapsto {\frak L}_f$ generated by
this correspondence using the letter ${\sf F}$.
Further, let us designate the power series of
$f(z)$ being the image of the element
$A \in {\frak L}$ under the mapping ${\sf F}: {\frak L} \,
\mapsto\, {\frak L}_f$ by $ {\sf F} [z| A]$.
The mapping ${\sf F}: {\frak L} \mapsto {\frak L}_f$ is obviously linear, i.e.
following relations take place
$${\sf F} [z| A + B] = {\sf F} [z| A] + {\sf F} [z| B] \,, \quad
{\sf F} [z| \lambda A] = \lambda {\sf F} [z| A] \,.$$

For each sequence $A\in {\frak L}$, it is possible to consider each of its
component $a_k, k \in {\Bbb N}_+$ as the appropriate projection of the infinite
ordered collection $A$. We shall write down this fact by means
$a_k = (A)_k $, $k \in {\Bbb N}_+$.

Let us introduce the binary commutative operation on ${\frak L}$
which we shall name
the {\it convolution} of sequences. We designate it
by the symbol $\circ$. This operation is determined as follows.
With any pair of sequences $A$, $B \in {\frak L}$, we associate the sequence
$A \circ B$ having components
$$(A\circ B)_k = \sum_{j = 0}^k a_j b_{k-j} \, \quad k=0,1,2... \,.$$
It is easy to see that the associativity property is fulfilled
for the introduced convolution operation applied for any three
elements $A, B, C \in {\frak L}$,
$$(A\circ B) \circ C = A \circ (B \circ C)\,.$$
It concerns also the distributivity property  relatively the addition,
$$(A + B) \circ C = A \circ C + B\circ C \,.$$
Besides, the following relation
$$(\lambda A) \circ B = \lambda (A \circ B)$$
takes place for any pair $A, B\in {\frak L} $
and for any $\lambda \in {\Bbb C} $.
Thus, these equalities together with the commutative property of the
convolution operation permits to conclude that the linear manifold
${\frak L}$ equipped with the "multiplication" $\circ $ turns into
the commutative algebra which we shall designate by
the same symbol ${\frak L}$.

In the algebra ${\frak L}$, there exists the unity $E$ which is represented
by the collection $E = \langle 1, 0, 0...\rangle$ since  for any
$A\in {\frak L}$, the following relation takes place
$$ A\circ E = E \circ A = A \,. $$

Let us notice that the subalgebra ${\frak L}_0 =
\{A: (A)_0 = 0 \}$ of the algebra ${\frak L} $ is its ideal,
i.e. it takes place $A\circ B \in {\frak L}_0 $
for any $A\in {\frak L}_0$ and for any $B \in {\frak L}$.

There exists the inverse element $A^{-1}$ for any element
$A \not\in {\frak L}_0 $ which has the property formulated by the following
way
$$A^{-1} \circ A = A \circ A^{-1} = E \,. $$
One may find consecutively each its component by the equality system
$$(A^{-1}) _0 = a_0^{-1} \,$$
$$(A^{-1}) _n = - a_0^{-1} \sum_ {k=1}^{n-1} (A^{-1})_{n-k} a_k \,
\quad n=1, 2, ... \,.$$

Let us introduce the following reduced
designation of powers of any element $A \in {\frak L} $. We shall write
$$A^0_\circ = E, \quad A_\circ^1 = A, \quad A\circ A = A_\circ^2 ,\quad ...
\quad,
A_\circ^n \circ A = A_\circ^{n+1}, \quad n = 0, 1, 2... \,. $$
Each component of the $n$th power of the element $A = \langle a_m;
m \in {\Bbb N}_+ \rangle$ is obtained by the formula
$$(A^n_\circ)_m = \sum_{k_1..., k_n \ge 0\atop
k_1 +... + k_n = m} a_{k_1}... a_{k_n} \, \quad m=1,2,... \,.\eqno (8)$$
Thus, it is obvious that if $A \in {\frak L}_0$ then the formula
$$\left (A_\circ^n\right)_m = \sum_{n > j_1..., j_n \ge 1,
\atop j_1 +... + j_n = m} a_{j_1}... a_{j_n} \eqno (9)$$
takes place. Therefore, the relation
$$\left (A_\circ^n\right)_m = 0 \eqno (10)$$
is fulfilled for all $m > n$.

The functional ${\sf F} [z| A]$ is multiplicative relatively the introduced
multiplication operation, i.e. the identity
$${\sf F} [z| A\circ B] = {\sf F} [z| A] \, {\sf F} [z| B]  \eqno (11)$$
takes place for any pair of collections $A$ and $B$.
It is ascertained by the following transformations
$${\sf F} [z| A \circ B] = \sum_{m=0}^\infty z^m (A \circ B)_m =
\phantom {AAAAAAAAA} $$
$$ = \sum_{m=0}^\infty z^m \sum_{k=0}^m a_k b_{m-k}
= \sum_{k=0}^\infty z^k a_k \sum_{m=k}^\infty z^{m-k} b_{m-k} = $$
$$\phantom {AAAAAAAAA} =
\left (\sum_{k=0}^\infty z^k a_k\right) \left (\sum_{m=0}^\infty
z^{m} b_{m} \right) = {\sf F} [z| A] {\sf F} [z| B] \,. $$
They are correct in the common convergence area of both series
${\sf F} [z| A] $ and $ {\sf F} [A| B] $.

On the algebra ${\frak L}$, it is possible to consider "analytical" functions
$f_\circ (A)$. Each of them is defined for every $A \in {\frak L} $ by means
of the corresponding "power" series
$$f_\circ (A) = \sum_{n=0}^\infty c_n A_\circ^n \, \quad c_n
\in {\Bbb C}, \quad n \in {\Bbb N}_+ \,. \eqno (12) $$
Naturally, it has the sense under the condition of the componentwise
convergence. In view of Eq.(10), such serieses are finite for each
individual component of the function value $f_\circ (A)$
if the element $A$ is chosen in ${\frak L}_0$, i.e.
$$\left (f_\circ (A) \right)_n =
\sum_{k=0} ^n c_n \left (A_\circ^k\right)_n $$
Therefore, the series (12) converges by the definition.
In particular, we introduce the following function
$$(E - A)_\circ^{-1} = \sum_{n=0}^\infty A^n_\circ \,, \eqno (13)$$
by means of the formula (12), which is defined at $A \in {\frak L}_0$.
At last, for any function $f_\circ $ on elements of the algebra
${\frak L}$, the following formula is correct
$${\sf F} [z| f_\circ (A)] = f
({\sf F} [z|A]) \,. \eqno (14) $$
It is due to linear and multiplicative properties  of the functional
${\sf F} [z| \cdot]$.
Here, the function $f_\circ (\cdot)$ on ${\frak
L}$ in the lefthand side is determined by means of Eq.(12) and
the corresponding analytic function on ${\Bbb Z}$ determined by Eq.(7)
in the righthand  side is designated by $f(\cdot)$.

\begin{center}
{\bf 3. Moments of the random variable ${\bf {\tilde J}}$}
\end{center}

At first, let us consider the problem about the power expansion
on $\lambda$ of the generating function
$Q(\lambda) = {\sf E}e^{- \lambda {\tilde J}}$ of the random variable
$${\tilde J} = \int\limits^T_0 \left|\,{\tilde \zeta}(t)\,\right|^2 dt\,.$$
This function is evaluated by the formula
$$Q(\lambda) = \frac {4 r \nu \exp(\nu T)}{(r + \nu)^2 \exp (r T)
- (r - \nu)^2 \exp (- rT)}\eqno (15)$$
on the basis of Eq.(4). Here, $r = \sqrt{\nu^2 + 2\lambda \sigma} = \nu q$,
$$q = \left(1 + \frac{2\lambda
\sigma}{\nu ^2}\right)^{1/2}\,. \eqno(16)$$
We represent the function $Q(\lambda)$ in the form
$$Q(\lambda) = e^{T}G^{-1}(\lambda)$$
where
$$G(\lambda) =
Q^{-1}(\lambda) e^T = \frac 1{4q}\left[(1+q)^2 e^{qT} - (q-1)^2
e^{-qT} \right]\,.$$
Here, we have introduced the "dimensionless"  parameter $\nu T$ instead of
$T$ which we designated hereafter by the same letter $T$ if it will not cause
a misunderstanding.

At first, let us decompose the function $G(\lambda )$ in the series on $q$
powers. We have
$$G(\lambda) =
(4q)^{-1}\left[(1+q)^2 \sum_{n=0}^{\infty}\frac {(qT)^n}{n!} -
(q-1)^2 \sum_{n=0}^{\infty}(-1)^n\frac {(qT)^n}{n!} \right] =$$
$$= (4q)^{-1} \left[(1 + q^2)\sum_{n=0}^{\infty} \frac
{(qT)^n}{n!} \left(1 - (-1)^n\right) + 2q \sum_{n=0}^{\infty}
\frac {(qT)^n}{n!} \left(1 + (-1)^n\right)\right] =$$

$$= (2q)^{-1} \left[(1  + q^2)\sum_{n=0}^{\infty} \frac
{(qT)^{2n+1}} {(2n+1)!} + 2q \sum_{n=0}^{\infty} \frac {(qT)^{2n}}
{(2n)!}\right] =$$

$$= \sum_{n=0}^{\infty} \frac {(qT)^{2n}} {(2n)!} \left[1 + \frac
T{(2n + 1)}\left(1 + \frac {\lambda \sigma}{\nu^2}\right)
\right]\,.$$
Thus,
$$G(\lambda) = \sum_{n=0}^{\infty}\frac
{T^{2n}}{(2n)!}\,q^{2n} \left[1 + \frac T{(2n + 1)}\left(1 + \frac
{\lambda \sigma}{\nu^2}\right) \right]\,. \eqno (17)$$
Now, let us substitute the expression (16) of the variable  $q = (1 +
z)^{1/2}$, $z = 2\lambda \sigma/\nu^2$ into the formula (17),
$$G(\lambda) =
\sum_{n=0}^{\infty}\,\frac {T^{2n}}{(2n)!}\left(1 + z \right)^n
\left[1 + \frac T{(2n + 1)}\left(1  + \frac z2\right)\right]\,.$$
Then we use the binomial formula. As a result, we obtain
$$G(\lambda) = \sum_{n=0}^{\infty} \frac
{T^{2n}}{(2n)!}\left[\sum_{m=0}^{n} {n \choose m} z^m\right]
\left[1 + \frac T{(2n + 1)}\left(1  + \frac z2\right)\right] = $$
$$ = \sum_{m=0}^{\infty} \frac {z^m}{m!}(u_m +  v_m) + \frac 12
\sum_{m=0}^{\infty} \frac {z^{m+1}}{m!} v_m = \sum_{m=0}^{\infty}
\frac {z^m}{m!}\left[u_m +  \frac m2 v_{m-1} + v_m\right]\,,$$
where
$$u_m = \sum_{n=m}^{\infty}\frac {T^{2n}}{(2n)!}\,\frac
{n!}{(n-m)!}\quad m \in {\Bbb N}_+\,, \eqno (18)$$
$$v_m =
\sum_{n=m}^{\infty}\frac {T^{2n+1}}{(2n+1)!}\,\frac
{n!}{(n-m)!}\quad m \in {\Bbb N}_+\,. \eqno (19)$$

Consider the sequence $W = \langle w_m; m \in {\Bbb N}\rangle$
with components $w_0 = 0$ and
$$w_m = \frac 1 {m!} e^{-T}(u_m  +
m v_{m-1}/2 + v_m)\,,\quad m \in {\Bbb N}\,.\eqno (20)$$
One may note that $e^{-T}(u_0 + v_0) =1$. Therefore, the expansion of the
function $G(\lambda)$ in the series on $\lambda $ powers is
presented in the following form
$$G(\lambda) = e^T \left[1 +
\sum_{m=1}^{\infty} z^m w_m\right] = e^T\left(1 + {\sf F}[z|
W]\right)\,,\quad z = \frac {2 \lambda \sigma }{\nu ^2}\,. \eqno
(21)$$

Since $(W)_0 = 0$, the sequence $W$ is the element of the algebra
${\frak L}_0$. Therefore, having written down the decomposition (21)
in the form
$$G(\lambda) = e^T \left[1 - {\sf F}[z| - W]\right) \eqno(22)$$
and using  firstly Eq.(22) and secondly Eq.(13), we
obtain the following expansion of the function
$G^{-1}(\lambda)$ on $\lambda$ powers,
$$G^{-1}(\lambda) = e^{-T} \left(1 - {\sf F}[z|
- W]\right)^{-1} = e^{-T}{\sf F}\left[z|X\right]\,,$$
where $X =
(E + W)^{-1} = \langle x_n; n \in {\Bbb N}_+\rangle$, $x_0 = 1$,
$$x_n = \sum_{l=1}^n (-1)^l\left(W_{\circ}^l\right)_n\,, \quad n
\in {\Bbb N}\,. \eqno (23)$$
Thus, we find
$$Q(\lambda) = e^TG^{-1}(\lambda) =
{\sf F}[z| X]\eqno (24)$$
substituting the expansion (24) in the expression of the function
$Q(\lambda)$. 
Formula (24), in the combination with definition of the
variable $z$, permits to us to be convinced that the
following statement is valid.

Theorem\ 1.\ {\sl Moments $M_n$, $n \in {\Bbb N}$
of the random variable
${\tilde J}$ are presented by the formula}
$$M_n = (-1)^n n!
\left(\frac {2 \sigma }{\nu ^2}\right)^n x_n\,,\quad n \in {\Bbb
N} \eqno (25)$$
{\sl where
coefficients $x_n$ are determined by formulas}
(18) - (20), (23).

$\square$ The formula (25) follows directly from the definition of
moments $M_n $ on the basis of their generated function
$Q(\lambda)$. Let us use the expansion  (24) of $Q(\lambda)$.
Then, for any $n \in {\Bbb N}$, we have
$$M_n = (-1)^n
\left(\frac {\partial^n Q(\lambda )}{\partial \lambda^n }
\right)_{\lambda = 0} = \phantom {AAAAAAAAAAAAAAAAAAA}$$
$$\phantom {AAAAAAAAAA}= (-1)^n \left[\frac {\partial^n {\sf F}[z|
X]} {\partial z^n }\right]_{z=0} \left(\frac
{dz}{d\lambda}\right)^n = (-1)^n n! \left(\frac {2 \sigma }{\nu
^2}\right)^n x_n. \quad \blacksquare$$
\smallskip

We note that each coefficient $x_n$ has the sign $(-1)^n$, $n = 1, 2,...$
due to the positivity of moments $M_n$.

\begin{center}
{\bf 4.\ Estimations of moments}
\end{center}

For the solution of the accuracy estimation problem of Mandel's distribution
approximations, it is necessary to find some {\it a priori} estimations of
moments $M_n$ of the random variable $\tilde J$.
This section is devoted to the obtaining of such estimations.
We divide the obtaining in some simple steps.

L e m m a\ 1.\ {\sl At $\alpha  \in [0, 1/2]$, the following inequality
takes place}
$$\ln (1 - \alpha ) \ge - 2\alpha\,. \eqno (26)$$

$\square$\ At $\alpha = 0$, Eq.(26) turns into the exact equality. At
$\alpha \in [0, 1/2]$, the inequality
$$ - \frac 1{1 - \alpha } \ge -2$$
of derivatives of Eq.(26) both parts is valid.
We obtain Eq.(26) integrating last inequality using the equality
condition at $\alpha = 0$.\ $\blacksquare$
\smallskip

L e m m a\ 2.\ {\sl The inequality
$$\frac 1{(2n)!} \le \frac {e n}{2^{2n} (n!)^2} \eqno (27)$$
takes place.}

$\square$\ Following identical transformations are valid
$$(2n)! = 2^n
n!(2n-1)!! = 2^{2n}(n!)^2 \prod _{l=1}^n \left (1-\frac
1{2l}\right) =$$
$$ = 2^{2n}(n!)^2 \exp \left(\sum_{l=1}^n \ln
(1-(2l)^{-1})\right)\,. \eqno (28)$$
Let us use the inequality (26) for the estimation of the
logarithm from below,
$$\ln
(1-(2l)^{-1}) \ge - l^{-1}\,, \quad l = 1, 2, ..., n\,. \eqno (29)$$
We give upper estimation of the expression
$\displaystyle \sum\limits_{l=1}^{n} 1/l$
considering it as the integral sum of the
function $\varphi (\alpha ) = \alpha^{-1}$,
$$\sum\limits_{l=1}^{n} \frac 1l  < 1 + \int\limits_1^n \frac
{d\alpha }\alpha = 1 + \ln\,n\,. \eqno (30)$$
The inequality follows from the low estimation of the integral
in the righthand side by the rectangular method taking into account that
the function $\varphi (\alpha)$ is decreasing at $\alpha > 0$.

Now, applying inequalities (29) and (30) for the low estimation of
the righthand side of Eq.(28), we find
$$(2n)! > \frac {2^{2n}(n!)^2}{en}\,.\quad \blacksquare$$
\smallskip

L e m m a\ 3.\ {\sl The following upper estimation of coefficients
$u_m$, $m = 1, 2, ...$ is valid,}
$$u_m  \le
e\frac{\left(T/2\right)^{2m}}{(m-1)!}\,{\rm I}_0 (T) \eqno (31)$$
{where
$${\rm I}_0 (T) = \sum_{n=0}^\infty \frac {(T/2)^{2n}}{(n!)^2}$$
is the zero order Bessel function on the imaginary variable.}

$\square$\ Using the inequality (27), from the definition formula (18),
we have for any $m \in {\Bbb N}$ that
$$u_m =
\sum_{n=m}^{\infty}\frac {T^{2n}}{(2n)!}\,\frac {n!}{(n-m)!} \le e
\sum_{n=m}^{\infty}\frac {(T/2)^{2n}}{(n-1)!(n - m)!} \le $$
$$\le
e\frac{\left(T/2\right)^{2m}}{(m-1)!} \sum_{n=m}^{\infty}\frac
{(T/2)^{2(n-m)}}{[(n - m)!]^2} =
e\frac{\left(T/2\right)^{2m}}{(m-1)!}\,{\rm I}_0 (T)\,, $$
We apply here the elementary inequality $(n-1)!
\ge (m -1)! (n - m)!$ which takes place at $n \ge m$.\ $\blacksquare$
\smallskip

C o r o l l a r y.\ {\sl From Eq.}(31) {\sl it follows
the upper estimation}
$$v_m \le \frac {T}{2 m +1}u_m
= \frac e{m!}\left(T/2\right)^{2m+1}\, {\rm I}_0 (T) \eqno (32)$$
{\sl of coefficients $v_m$, $m=1, 2, ...$ since
$(2l + 1)! \ge (2 m +1)(2l)!$ for all $l \ge m$\,.}
\smallskip

Summarizing, it is possible to assert on the basis of Eqs.(20),
(31),(32) that the following lemma takes place.

L e m m a\ 4.\ {\sl For coefficients $w_m$, the following estimation is valid}
$$w_m < \frac {e}{m!(m-1)!}\left(T/2\right)^{2m} I_0 (T)\varphi(m)\,,
\eqno (33)$$ 
$$\varphi (m) = \left[1 + \frac T{2m} + \frac mT\right]\,.$$
\smallskip

L e m m a\ 5.\ {\sl For modules of coefficients $x_m$, 
at $m > 1$, following estimations take place
$$|x_m| < \frac {e \psi (T) I_0 (T)}{e
\psi (T) I_0(T) - T}\left(eT\psi (T)I_0 (T)/4\right)^m \eqno (34)$$ 
where the function $\psi (T)$ is determined by the equality
$$\psi (T) \equiv TI_1(2) + \frac{T^2}2(I_0(2) - 1) + I_0(2)\eqno (35)$$
and
$$I_1 (T) = \frac {dT}{dT} = \sum_{m=1}^\infty\frac {(T/2)^{2m-1}}{m! (m-1)!}$$
is the Bessel fubction of first order.}

$\square$\ At $m > 1$, the formula (23) rewrites as follows
$$x_m = - w_m + \sum_{k=1}^{m-1} (-1)^{k-1} \sum_{l_1, ..., l_k> 0\atop
l_1 + ... + l_k < m} w_{m - l_1 - ... - l_k} \prod_{j=1}^k
w_{l_j}\,. \eqno (36)$$
Then, using the estimation (33), we have
$$|x_m| \le |w_m| + \sum_{k=1}^{m-1} \sum_{l_1, ..., l_k > 0\atop
l_1 + ... + l_{k} < m} |w_{m - l_1 - ... - l_k}| \prod_{j=1}^k
|w_{l_j}| \le \phantom{AAAAAAAAA}$$
$$\phantom{AAAAAAAAA}\le
(T/2)^{2m}\sum_{k=1}^m\sum_{l_1, ..., l_k
> 0\atop l_1 + ... + l_{k} = m} \prod_{j=1}^k \frac {eI_0(T)}{l_j! (l_j
-1)!}\varphi(l_j)\le$$ 
$$\le (T/2)^{2m}\sum_{k=1}^m \prod_{j=1}^k
\sum_{l = 1}^m \frac {eI_0 (T)}{l! (l -1)!}\varphi(l)\le
\phantom{AAAAAAAAA}$$
$$\phantom{AAAAAAAAA} \le
(T/2)^{2m}\sum_{k=1}^m (eI_0(T))^k\left(\sum_{l = 1}^\infty \frac
{\varphi(l)}{l! (l -1)!}\right)^k\,.$$

According to the definition of Bessel's functions on the imaginary variable,
the sum in brackets in the last expression is equal to
$$\sum_{l = 1}^\infty \frac {\varphi(l)}{l! (l -1)!} =
I_1(2) + T\left(I_0(2) - 1\right)/2  + I_0 (2)/T \equiv \frac {\psi (T)}T$$
Then, taking in mind of the
above obtained estimation and the inequality $e\psi (T) > 1$, we find
$$|x_m| < (T/2)^{2m}\sum_{k=1}^m (eI_0(T))^k\left(\psi(T)/T\right)^k =
\frac {e \psi (T)I_0(T)}{e \psi (T) I_0 (T) - T}
\left(eT\psi (T)I_0(T)/4\right)^m\,. \quad
\blacksquare$$

C o r o l l a r y.\  {\sl Basing on Theorem 1, from obtained
estimations (34), we find following estimations of moments $M_n$,}
$$M_n = n! \left(\frac {2
\sigma }{\nu ^2}\right)^n |x_n| < \frac {e \psi (T)I_0 (T)}{e \psi (T)
I_0 (T) - T}\, n! \left(\frac {e \sigma T \psi(T) I_0 (T)}{2\nu ^2}\right)^n\,.$$

R e m a r k.\ At $T \to 0$, following asymptotic formulas take place
$$u_m = \left[T^{2m}/(2m)!\right](1 + o(1))\,,
v_m =\left[T^{2m+1}/(2m+1)!\right](1 + o(1))\,.$$
Then
$$w_m = \left[T^{2m -1}/ 2(m-1)!(2m-1)!\right](1 + o(1))$$
and, according to the
formula (36), the basic contribution in $x_m$ gives
the term with $k=1$ at $T \to 0$, i.e.
$x_m = \left[(-1)^m (T/2)^m\right](1 +
o(1))$. Therefore, since $\psi (0) =$ const, the obtained estimation
(34) is asymptotically exact at $T \to 0$.

\begin{center}
{\bf  5.\ Approximations of the Mandel distribution  \\ and
estimations of their accuracy}
\end{center}

Let us consider the Mandel distribution (1).
We present it as the expansion on moments
$$P_n = \frac 1{n!} {\sf
E}\sum_{l=0}^\infty \frac {(-1)^l}{l!} {\tilde J}^{l+n} = \frac
1{n!} \sum_{l=0}^\infty \frac {(-1)^l}{l!} M_{l+n}\,,$$
or, on the basis of the representation of moments (25), it gives
$$P_n = \frac
{(-1)^n}{n!}\sum_{l=0}^\infty \frac {(n+l)!}{l!} \left(\frac {2
\sigma}{\nu^2}\right)^{n+l} x_{l+n}\,. \eqno (37)$$
We determine the sequence of approximations $P_n^{(N)}$, $N=1, 2,
...$ of the probability distribution which present it with the accuracy up
to the $N$th power of the parameter $(\sigma T/\nu^2)$
(see Remark of Lemma 5) when $n \le N$,
$$P_n^{(N)} = \frac {(-1)^n}{n!}\sum_{l =
0}^{N-n}\frac {(n+l)!}{l!} \left(\frac {2
\sigma}{\nu^2}\right)^{n+l} x_{l+n}\,. \eqno (38)$$

Further, we estimate the deviation of the $(N-1)$th approximation
from the exact distribution (1). On the basis Eq.(37), we have
$$|P_n - P_n^{(N-1)}|
\le \frac 1{n!} \sum_{l = N-n}^\infty \frac {(n+l)!}{l!}
\left(\frac {2 \sigma}{\nu^2}\right)^{n+l} |x_{l+n}| =$$
$$ = \frac 1{n!}\sum_{l = N}^\infty \frac {l!}{(l-n)!} \left(\frac {2
\sigma}{\nu^2}\right)^{l}|x_{l}|\,.$$
Then we use the estimation (34),
$$|P_n - P_n^{(N-1)}| \le \frac 1{n!}\, \frac {e \psi I_0 (T)}{e \psi 
I_0 (T) - T}
\sum_{l = N}^\infty \frac {l!}{(l-n)!} \left(\frac {2
\sigma}{\nu^2}\right)^{l} \frac {\left(eT\psi I_0 (T)\right)^l}
{2^{2l}}\,, \eqno(39)$$
where the variable of the function $\psi$
hereinafter is not pointed out explicitly. Let us give separately
the estimation of the series
$$R_N(\zeta, n) =
\sum\limits^\infty_{l = N} \frac {l!}{(l-n)!} \zeta^l$$
for positive values of $\zeta$ and for $n \ge N$.
The following transformations are valid
$$R_N(\zeta, n) = \zeta^n
\sum\limits^\infty_{l = N} \frac {l!}{(l-n)!} \zeta^{l-n} =
\zeta^n \frac {d^n}{d\zeta^n} \sum_{l = N}^\infty \zeta^l =
\zeta^n \frac {d^n}{d\zeta^n} \frac {\zeta^N}{1 - \zeta} = $$
$$= \zeta^n \sum_{l=0}^n {n \choose l}\frac {N!}{(N - l)!} \zeta^{N-l}
\frac {(n - l)!}{(1 - \zeta)^{1 + n - l}} = \frac {\zeta^N}{1 -
\zeta} \sum_{l=0}^n n! {N \choose l}\left(\frac {\zeta} {1
- \zeta}\right)^{n - l} \le $$

$$ \le n! \frac {(2\zeta)^N }{1 - \zeta} \sum_{l=0}^n 
\left(\frac \zeta {1 - \zeta}\right)^{n - l} \le 
n!\frac {(2\zeta)^N}{1 - 2\zeta}\,.$$
Applying the obtained estimation to the inequality (39), we come
to the following statement.

T h e o r e m\ 2.\ {\sl The $N$th approximation $P^{(N)}_n$ of the
Mandel distribution presents this distribution with
the guaranteed accuracy determined by the inequality
$$|P_n - P_n^{(N)}| \le  \left(\frac {e \psi I_0 (T)}{e \psi I_0(T) - T}\right)
\left(\frac {(2\zeta)^{N+1}}{1 - 2\zeta}\right) \eqno (40)$$
if}
$$\zeta = \frac {e \sigma T \psi}{2 \nu^2}I_0 (T) < \frac 12\,.$$
\smallskip

In order to obtain the effective algorithm of consecutive calculation of
$P_n^{(N)}$, it is necessary to solve the last problem. It
consists of the creation of the method which permits to find consecutively all
components of the sequence $X$.

L e m m a\ 6.\ {\sl The following formula takes place
$$\left(\frac {\partial^m}
{\partial \alpha^m } \exp(\pm\alpha^{1/2} T)\right)_{\alpha =1} \
=\ m!e^{\pm T}R_m^{\pm}(T) \eqno (41)$$
where polynoms $R_m^\pm (T)$ of the $m$ degree on the variable $T$
are determined by the recurrent relation}
$$R_{m+1}^\pm(T) = \pm \frac T {2(m+1)}
\sum_{l=0}^m (-1)^l \frac {(2l)!} {2^{2l}(l!)^2} R_{m-l}^\pm
(T)\,,\quad m \in {\Bbb N}_+  \eqno (42)$$
{\sl and by the condition at $m=0$}
$$R_0^\pm (T) = 1\,. \eqno (43)$$

$\square$\  The proof is realized by the induction on $m$.
At $m=0$, we obtain (42). We shall build the induction step.
Let Eq.(41) takes place at the given value $m \in {\Bbb N}_+$.
Then we shall introduce the function $R_{m+1}(T)$ according to
$$(m+1)!e^{\pm T} R_{m+1} (T) =
\left(\frac {\partial^{m+1}} {\partial \alpha^{m+1} }
\exp\left(\pm\alpha^{1/2} T\right)\right)_{\alpha =1} =$$
$$ = \pm
\left(\frac {\partial^{m}} {\partial \alpha^{m}}\left[\frac
T{2\alpha^{1/2}} \exp(\pm\alpha^{1/2} T)\right]\right)_{\alpha =1}\,.$$
Using the induction assumption and the formula
$$(\alpha^{-1/2})^{(l)}\ =\ (-1)^l\displaystyle \frac {(2l)!}{2^{2l} l!}
\alpha^{-(2l+1)/2}\,,$$
we accomplish the $m$-fold differentiation,
$$(m+1)!e^{\pm T} R_{m+1} (T) = \phantom
{AAAAAAAAAAAAAAAAAAAAAAAAAAAA}$$
$$= \pm \frac T2 \sum_{l=0}^m{m \choose l} \left(\frac {\partial^{m-l}}
{\partial \alpha^{m-l}}\exp\left( \pm \alpha^{-1/2} T\right)
\right)_{\alpha =1} \left(\frac {d^{l}} {d
\alpha^{l}}\alpha^{-1/2} \right)_{\alpha =1} = $$
$$\phantom
{AAAAAAAAAAAAAAAA}= \pm m! T
 \frac {e^{\pm T}}2\sum_{l=0}^m \frac {(-1)^l}{l!} R_{m-l}^\pm (T)
\frac {(2l)!}{2^{2l} l!}\,. $$
Whence formulas (41), (42) follow and although the fact that the function
$R_{m+1}(T)$ is the polynom of the $m + 1$ degree is proved.\
$\blacksquare$
\smallskip

T h e o r e m\ 3.\ {\sl Components of sequences
$U  = \langle u_m; m \in {\Bbb N}\rangle$ and $V  = \langle v_m; m \in
{\Bbb N}\rangle$ are represented in the form
$$u_m = \frac 12 m!
\left(e^T R_m^+ (T) + e^{-T} R_m^-(T)\right)\,,\eqno (44)$$
$$v_m = \frac 1T (m+1)!\left(e^T R_{m+1}^+ (T) + e^{-T} R_{m+1}^-
(T)\right) \eqno (45)$$
where polynoms $R_m^{\pm}(T)$ are calculated by the formula (42).}

$\square$\ Using Eq.(41),
from the definition of coefficients $u_m$, we have
$$u_m = \sum_{n = m}^\infty \frac {T^{2n}}{(2n)!} \frac
{n!}{(n-m)!} = \left(\frac {\partial^m}{\partial \alpha^m} 
\sum_{n = m}^\infty \frac {\alpha^nT^{2n}}{(2n)!}\right)_{\alpha  =1}  =$$
$$ = \left(\frac
{\partial^m}{\partial \alpha^m} {\rm ch}(\alpha^{1/2} T)
\right)_{\alpha  =1}= \frac 12 \left(\frac {\partial^m}{\partial
\alpha^m}\left(\exp(\alpha^{1/2} T) + \exp (-
\alpha^{1/2}T)\right)\right)_{\alpha  =1}= $$
$$= \frac 12 m! \left(e^T R_m^+ (T) + e^{-T} R_m^-(T)\right)$$

On the basis of the formula (41), we have although
$$\left(\frac {\partial^m}
{\partial \alpha^m }\left[\alpha^{- 1/2}\exp\left( \pm
\alpha^{1/2} T \right)\right]\right)_{\alpha =1} = $$
$$ =
\sum_{l=0}^m {m \choose l} \left(\frac {\partial^{m-l}} {\partial
\alpha^{m - l}} \exp\left(\pm \alpha^{1/2} T
\right)\right)_{\alpha =1} \left(\frac {\partial^{l}} {\partial
\alpha^{l}} \alpha^{- 1/2}\right)_{\alpha =1} =$$
$$ = m! e^{\pm
T}\sum_{l=0}^m (-1)^l R^{\pm}_{m-l}(T) \frac {(2l)!}{2^{2l}(l!)^2} = \pm 
\frac 2T (m+1)! e^{\pm T}R^{\pm}_{m+1} (T)\,.$$
Further, on the basis of the obtained formula, from the definition of
coefficients $v_m$, we have
$$v_m = \sum_{n = m}^\infty \frac {T^{2n+1}}{(2n+1)!}
\frac {n!}{(n-m)!} = \left(\frac {\partial^m}{\partial \alpha^m}
\sum_{n = m}^\infty \frac {\alpha^nT^{2n+1}}{(2n+1)!} 
\right)_{\alpha  =1}  =$$
$$ = \left(\frac {\partial^m}{\partial \alpha^m} \frac {{\rm
sh}(\alpha^{1/2}T)}{\alpha^{1/2}}\right)_{\alpha  =1} = \frac 12
\left(\frac {\partial^m}{\partial \alpha^m}\left[\frac
{\exp(\alpha^{1/2} T)}{\alpha^{1/2}} - \frac {\exp (-
\alpha^{1/2}T)}{\alpha^{1/2}}\right]\right)_{\alpha  =1} =$$
$$ = \frac 1T (m+1)! \left(e^T R^+_{m+1} (T) + e^{- T}R_{m+1}^- (T)
\right)\,. \ \ \blacksquare$$
\smallskip

C o r o l l a r y.\ {\sl Taking into account Eqs.(44),(45) and
Eq.(20), components of the sequence $W$ are represented as follows}
$$w_m = \frac 12R^+_m (T)\left(1 + \frac m T \right) +
\frac {m+1}T R^+_{m+1} (T) + \phantom{AAAAAAAAAAAAAAA} $$
$$\phantom{AAAA} + e^{- 2T}\left[\frac 12 R^-_m (T)\left(1 + \frac
m T \right) + \frac {m+1}T R^-_{m+1} (T)\right]\,, \quad m \in
{\Bbb N}\,. \eqno (46)$$
\smallskip

\begin{center}
{\bf 6. Approximate formulas}
\end{center}

Thus, Theorem 3 permits to calculate successively
all approximations $P^{(N)}_n$ of probabilities $P_n$ using formulas (46). 
In this section,
we shall give  some first approximations of the probability
distribution $P_n$ by means of this algorithm.

At first, we find the explicit form of polynoms
$R_m^\pm (T)$, $m =0, 1, 2, 3, 4$,
$$R_0^{\pm}(T) = 1\,, \quad  R_1^{\pm}(T) = \pm T/2\,,$$
$$R_2^\pm (T)  = \pm (T/2^2) \left(R_1^\pm (T) - 1/2 \right) =
(T/2^3) (T \mp 1),$$
$$R_3^\pm (T) = \pm (T/2 \cdot 3)
\left(R^\pm_2 (T) - R_1^\pm(T)/2 + 3/8 \right)
 = \pm (T/2^4 \cdot 3)\left(T^2 \mp 3 T + 3 \right),$$
$$R_4^\pm (T) = \pm (T/2^3)\left(R^\pm_3 (T) - R_2^\pm(T)/2 +
3R_1^\pm (T)/8  - 5/16\right) =$$
$$ = (T/2^7\cdot3)\left(T^3 \mp 6T^2 + 15 T \mp 15 \right)\,.$$
On their basis, we calculate components $w_m$, $m=1, 2, 3$.
$$w_1 =
\frac T2\,, \quad w_2 =  \frac 1{2^4}\left(2T^2 - 2T + 1 -
e^{-2T}\right)\,, \eqno (47)$$
$$w_3 =  \frac 1{3 \cdot
2^5}\left(2T^3 - 6T^2 + 9T  - 6 + 3(T+2) e^{-2T}\right)\,. \eqno (48)$$
At last, we find components $x_m$, $m=1, 2, 3$, by the formula (36),
$$x_0 = 1\,, \quad x_1 = - w_1\,, \quad x_2 = - w_2 +
w_1^2\,, \quad x_3 = - w_3 + 2 w_1 w_2 - w_1^3$$
or after the substitution (46), (47), we have
$$x_1 = - \frac T2, \quad x_2 = \frac
1{2^4}\left(2 T^2 + 2T -1 + e^{- 2T} \right)\,,$$
$$x_3 = - \frac
1{3 \cdot 2^5}\left[2 T^3 + 6 T^2 + 3T - 6 + 3(3T+ 2) e^{-
2T}\right]\,.$$

Explicit expressions of components $x_m$, $m = 0,1,2,3$ permit to us
to write the approximated expressions of $n$-photon registration
probabilities for $n = 0, 1, 2, 3$ up to the third order ($N =
0, 1, 2, 3$).
All probabilities $P_n^{(N)}$ of some higher values $n > 3$
are equal to zero in this approximation. We result consequtively
those expressions for each order of the approximation.

For  $N=0$, we have that only the probability $P_0^{(0)} = 1$ differs from 
zero.

At $N =1$, we obtain
$$P_0^{(1)} = \sum_{l = 0, 1}
\left(\frac {2\sigma}{\nu^2}\right)^l x_l = 1 - \frac {\sigma
T}{\nu^2}\,, \quad P_1^{(1)} = \frac {\sigma T}{\nu^2}\,.$$

At $N=2$, we have correspondingly
$$P_0^{(2)} = \sum_{l = 0, 1, 2}
\left(\frac {2\sigma}{\nu^2}\right)^l x_l = 1 - \frac {\sigma
T}{\nu^2} + \left(\frac \sigma {2 \nu^2}\right)^2 (2 T^2 + 2T - 1
+ e^{- 2T})\,,$$
$$P_1^{(2)} = \frac {\sigma T}{\nu^2} - 2
\left(\frac \sigma {2 \nu^2}\right)^2 (2 T^2 + 2T - 1 + e^{- 2T})\,,$$
$$P_2^{(2)} = \left(\frac \sigma {2 \nu^2}\right)^2 (2
T^2 + 2T - 1 + e^{- 2T})\,.$$

At last, for $N=3$, we get following approximate formulas
$$P_0^{(3)} = \sum_{l = 0, 1,
2, 3} \left(\frac {2\sigma}{\nu^2}\right)^l x_l = $$
$$ = 1 -
\frac {\sigma T}{\nu^2} + \left(\frac \sigma {2 \nu^2}\right)^2 (2
T^2 + 2T - 1 + e^{- 2T}) - \phantom {AAAAAAAAAAA}$$
$$\phantom
{AAAAAAAAAAA} - \frac 23 \left(\frac \sigma {2 \nu^2}\right)^3
\left[2 T^3 + 6T^2 + 3T - 6 + 3(3T +2)e^{- 2T}\right]\,,$$
$$P_1^{(3)} = - \sum_{l = 0, 1, 2} (l+1) \left(\frac
{2\sigma}{\nu^2}\right)^{l+1} x_{l+1} = $$
$$ = \frac {\sigma
T}{\nu^2} - 2\left(\frac {\sigma}{2\nu^2}\right)^2 \left(2 T^2 +
2T -1 + e^{- 2T} \right) + \phantom {AAAAAAAAAAAAA}$$
$$\phantom
{AAAAAAAAA} + 2 \left(\frac {\sigma}{2\nu^2}\right)^3\left[2 T^3 +
6 T^2 + 3T - 6 + 3(3T+ 2) e^{- 2T}\right]\,,$$
$$P_2^{(3)} = \frac 12 \sum_{l=0, 1} (2 + l)(1 + l) \left(\frac
{2\sigma}{\nu^2}\right)^{l+2}x_{l+2} = $$
$$= \left(\frac
{\sigma}{2\nu^2}\right)^{2} \left(2 T^2 + 2T -1 + e^{- 2T} \right)
-  \phantom {AAAAAAAAAAAAAA}$$
$$\phantom {AAAAAAAAAAA} -
2\left(\frac {\sigma}{2\nu^2}\right)^{3} \left[2 T^3 + 6 T^2 + 3T
- 6 + 3(3T+ 2) e^{- 2T}\right]\,,$$
$$P_3^{(3)} =  \frac 23
\left(\frac {\sigma}{2\nu^2}\right)^3 \left[2 T^3 + 6 T^2 + 3T - 6
+ 3(3T+ 2) e^{- 2T}\right]\,.$$

\begin {center}
{\bf 7. Conclusion}
\end {center}

It is necessary to point out some lacks of the problem setting and its
solution in the given work.

Our research is based on the expansion (5) which is directly connected with
the expansion of the generating function (15) on $\lambda$ powers.
The convergence radius of last expansion is equal to the convergence
radius $r (T, \nu, \sigma) $ of the power series on $\lambda$ of the
function $Q (\lambda) $ which is meromorphic only. It has poles in the
complex plane $ \lambda \in { \Bbb C} $. It means that the distance to nearest
of these poles is equal to $r (T, \nu, \sigma) $.
It is clear that $r (T, \nu, \sigma) \to 0 $ when
parameters $T $ or $ \sigma $ increasing.
At the same time, such restriction of the convergence domain
of the expansion (5) is not caused by a physical reason,
i.e. the divergence of the series (5) is not connected
with the presence of anything qualitative change of the
 probability distribution $P_n $ behavior
when parameters $T, \nu, \sigma $ varying.
Therefore, it is desirable to get rid of this lack of the expansion (5).

Other lack is connected with the above mentioned one.
With irreversibility, we make a mistake when calculating the convergence
radius by the estimation of the residual series (6) basing
on a priori estimations of moments ${\sf E}{\tilde J}^m$, $m\in {\Bbb N}$.
Actually, only its low bound is obtained by this way and it is sufficiently
coarse as one may see from our work.

It is possible to consider mentioned lacks as ones generated by
our method when
the problem solving. The following lack concern with the unsuccessful problem
setting.

Let us pay attention that obtained formulas permit to calculate
the registration probability of everyone concrete photon number $n$.
However, they do not permit to calculate (to estimate)
the registration probability in the case when this number is indefinite,
i.e. in the case when it is not known exactly and, therefore, the number
$n$ is  the parameter of the problem.
It is connected with the fact that their analytical representation
becomes very tedious when the approximation order increasing.

\end{document}